\documentclass{PoS}

\usepackage{amsmath}

\title{Nucleon Structure with Domain Wall Fermions \\
       at \boldmath{$a$} = 0.084\,fm
         }

\ShortTitle{Nucleon Structure with Domain Wall Fermions at a = 0.084 fm}

\author{\speaker{S.~N.~Syritsyn}, 
        J.~D.~Bratt, 
        M.~F.~Lin, 
        H.~B.~Meyer, 
        J.~W.~Negele, 
        A.~V.~Pochinsky 
        and M.~Procura, \\
Center\hspace{-0.0725pc} for\hspace{-0.0725pc} Theoretical\hspace{-0.0725pc} Physics,\hspace{-0.0725pc} Massachusetts\hspace{-0.0725pc} Institute\hspace{-0.0725pc} of\hspace{-0.0725pc} Technology,\hspace{-0.0725pc} Cambridge,\hspace{-0.0725pc} MA\hspace{-0.0725pc} 02139,\hspace{-0.0725pc} USA \\
        E-mail: 
          \email{syritsyn@mit.edu},
          \email{jdbratt@mit.edu},
          \email{meifeng@mit.edu},
          \email{meyerh@mit.edu},
          \email{negele@mit.edu},
          \email{avp@mit.edu},
          \email{mprocura@mit.edu}
        }

\author{R.~G.~Edwards, K.~Orginos and D.~G.~Richards\\
        Thomas Jefferson National Accelerator Facility, Newport News, VA 23606, USA \\
        E-mail: 
          \email{edwards@jlab.org},
          \email{dgr@jlab.org},
          \email{kostas@jlab.org}
        }
\author{M.~Engelhardt\\
        Department of Physics, New Mexico State University, Las Cruces, NM 88003-8001, USA \\
        E-mail: \email{engel@physics.nmsu.edu}}
\author{G.~T.~Fleming\\
        Sloane Physics Laboratory, Yale University, New Haven, CT 06520, USA \\
        E-mail: \email{George.Fleming@Yale.edu}}
\author{Ph.~H\"agler and B.~Musch, \\
        Institut f\"ur Theoretische Physik T39, Physik-Department der TU M\"unchen,
        James-Franck-Strasse, D-85747 Garching, Germany \\
        E-mail: 
          \email{phaegler@ph.tum.de},
          \email{bernhard.musch@ph.tum.de}
        }
\author{D. B. Renner\\
        DESY Zeuthen, Theory Group, Platanenallee 6, D-15738 Zeuthen, Germany \\
        E-mail: \email{drenner@ifh.de}}
\author{W. Schroers\\
        Department of Physics, National Taiwan University, Taipei 10617, Taiwan \\
        E-mail: \email{Wolfram.Schroers@Field-theory.org}}

\abstract{
We present  initial calculations of nucleon matrix elements of twist-two operators 
 with 2+1 flavors of domain wall fermions at a lattice spacing 
a = 0.084 fm for pion masses down to 300 MeV.
We also compare the results with the domain wall calculations on a coarser lattice.
}

\FullConference{The XXVI International Symposium on Lattice Field Theory \\
                 July 14 - 19, 2008\\
                 Williamsburg, Virginia, USA}

\begin{document}

\section{Introduction}
The calculation of nucleon generalized form factors has been performed recently 
using a mixed action that combines the computationally economical Asqtad fermion action 
for sea quarks with the chirally symmetric domain wall action for
valence quarks\cite{milc-ff-papers,meifeng-john-talk}.

With the advance of algorithms \cite{rbc-rhmc} and computational facilities, 
use of the domain wall action for light sea quarks on large, fine lattices has now become possible.
Hence, using gauge configurations generated by the RBC and UKQCD collaborations \cite{rbc-ukqcd-gauge},
we investigate  nucleon structure in  fully unitary, chirally symmetric lattice QCD. 
Currently, lattices with two different lattice spacings are available,
$a=0.114\,\mathrm{fm}$ and $a=0.084\,\mathrm{fm}$, which we will refer to as \emph{coarse} and
\emph{fine}, respectively. 
The lowest pseudoscalar meson mass is $m_\pi\approx300\,\mathrm{MeV}$.

The next section describes the details of our calculation.
In the Sect.~\ref{sect:pion-corr}, we study the pion correlation functions to determine
the renormalization constant for the axial vector quark current, 
the pion decay constant and the pion mass, and use the pion decay constant and mass 
to set the scale of the fine lattices. 
The results for nucleon form factors 
are given in Sect.~\ref{sect:ff-results}, where we focus on the isovector flavor combination 
$u-d$ since it has no contributions from  disconnected diagrams. 
We also show the $Q^2$ dependence of the generalized form factors, although we still need to
determine their overall renormalization.
Conclusions follow in Sect.~\ref{sect:conclusions}.

\section{\label{sect:calc-details} 
  Calculation details}
The gauge configurations were generated by the RBC and UKQCD collaborations using the Iwasaki gauge
action with $N_f=2+1$ light and heavy dynamical Domain Wall fermions, as described in 
Ref.~\cite{rbc-ukqcd-gauge} and references therein.
The extent of the fifth dimension is $L_s=16$, which is large enough to keep the residual quark mass
below the bare quark mass as shown in  Tab.~\ref{tab:gauge-conf}. 
The spatial volume is $\approx(2.7\,\mathrm{fm})^3$ for both lattice
spacings.

\begin{figure}
\begin{minipage}{.49\textwidth}
\includegraphics[angle=-90,width=\textwidth]{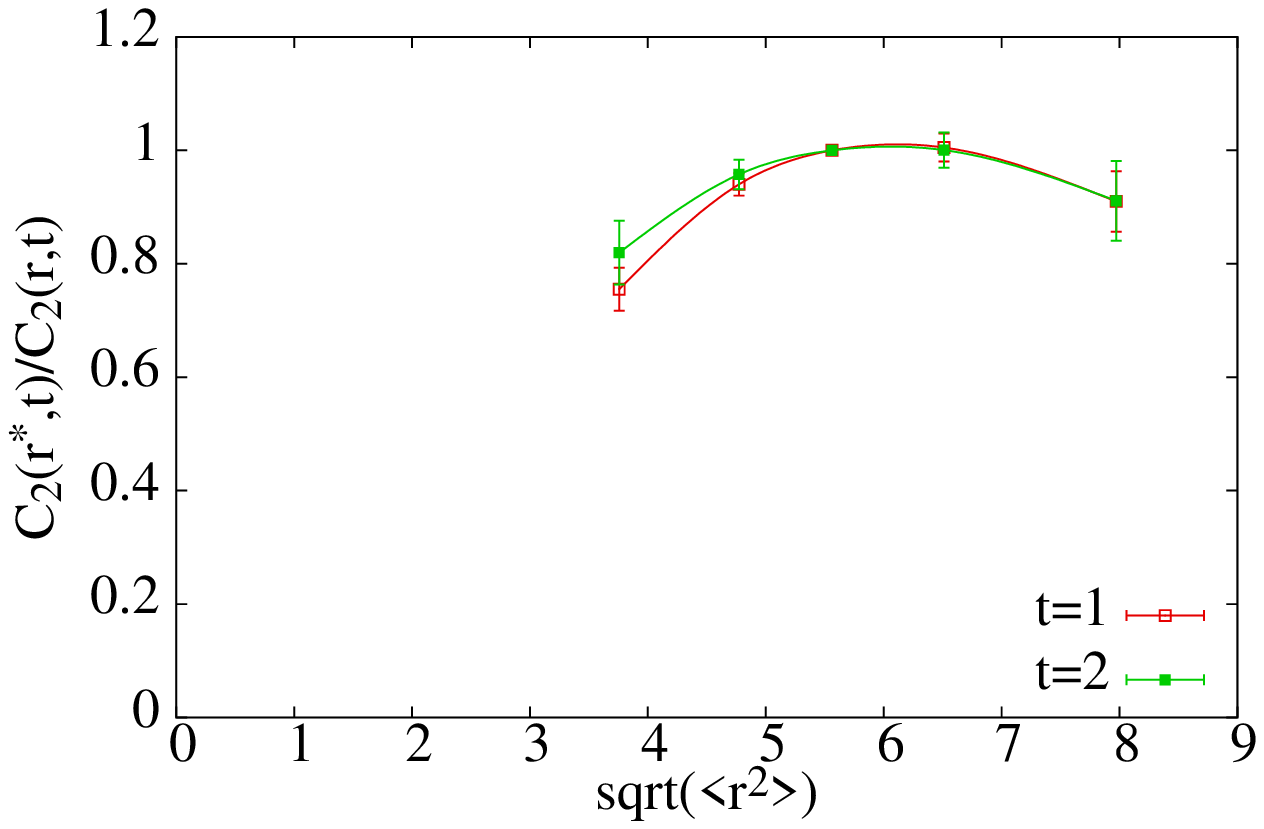}
\vspace{3pt}\\
\centerline{(a)}
\end{minipage}
\begin{minipage}{.49\textwidth}
\includegraphics[width=\textwidth]{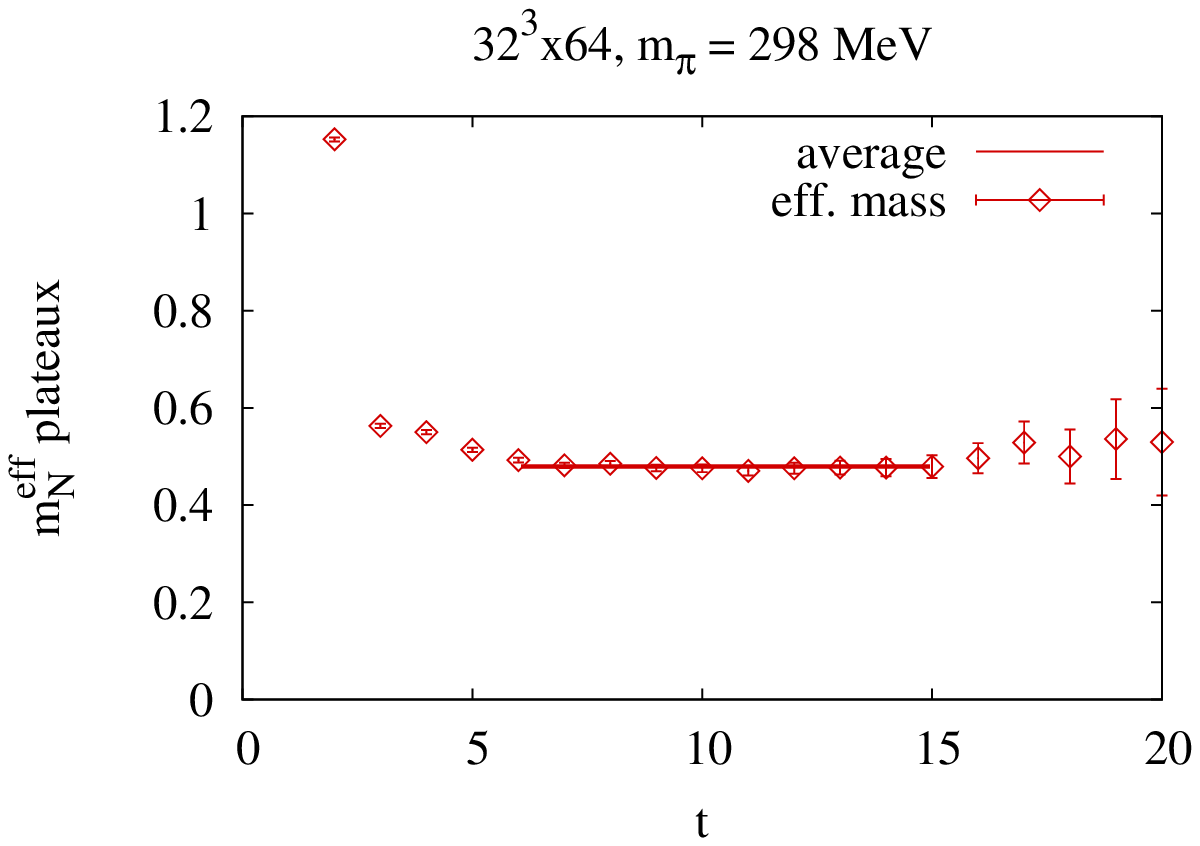}
\centerline{(b)}
\end{minipage}
\caption{\label{fig:meff-pltx} 
  Source is optimized studying the overlap with the ground state (a).
  The effective mass plateau starts at $t=6$ (b).}
\end{figure}

Before analyzing the gauge configurations, we undertook a systematic search for the optimal
source parameters that provide the best overlap with the nucleon ground state. 
As in Ref.~\cite{milc-ff-papers}, we use a gauge-invariant Gaussian-smeared quark source that
minimizes the excited state contamination to the
nucleon two-point correlation function. 
In addition, we apply APE smearing to the gauge field used to 
construct the sources to reduce the  large variation of the norm of the smeared sources due to the gauge
field noise. The optimization of the source overlap with the nucleon is shown in Fig.~\ref{fig:meff-pltx}a.
With the optimized source, the plateau for the effective nucleon mass starts as early as $t=6$
for the fine lattice (see Fig.~\ref{fig:meff-pltx}b) and $t=5$ for the coarse lattice.
This justifies our choice of the source-sink separation $T=9$ and $T=12$ for coarse
and fine lattices, respectively, which both correspond to physical separations 1.0 fm,
for the calculation of the nucleon three-point correlators.

As described in Ref.~\cite{meifeng-john-talk}, to increase the statistics, 
we use four nucleon sources separated by T = 16 and calculate the forward quark
propagators. 
The backward propagators are calculated for the sum of four nucleon  sinks on each lattice.
The cross-contributions between different sources and sinks average to zero due to 
the gauge invariance, provided there is no temporal link gauge fixing.
Similarly, four antinucleon sinks are also treated analogously to obtain a total of  
eight measurements per lattice, which have been verified to be independent by jackknife 
binning\cite{meifeng-john-talk}. 

For each source, we construct sinks with momenta $\vec{P^\prime}=(0,0,0)$ 
and $\vec{P^\prime}=(-1,0,0)$. 
Since the three-point correlators quickly become noisy with growing initial  and final state momenta, 
we have limited the source momenta to $\vec P^2 \le 4$ for the non-zero sink momentum
$\vec{P^\prime}$.

\begin{table}[ht]
\caption{\label{tab:gauge-conf}
  Dynamic DW fermion gauge configurations calculated by the RBC/UKQCD collaboration.
  The  total number of nucleon correlator measurements, $\#$, includes eight measurements per gauge field.
  }
\begin{center}
\begin{tabular}{ll|l|lll}\hline
 & $a\, [fm]$ & \# & $am_l/am_h$ & $am_{res}\times 10^3$ & $m_\pi [MeV]$ \\
\hline
$24^3\cdot64$  & $0.114$ & 
3208 & $0.005/0.04$ & $3.15(1)$ & $329(5)$ \\
\hline
$32^3\cdot64$  & $0.084$  & 
1568  & $0.008/0.03$ & $0.668(3)$ &  $406(7)$ \\
& &  
4208  & $0.006/0.03$ & $0.663(2)$ &  $356(6)$ \\
& & 
2392  & $0.004/0.03$ & $0.665(3)$ &  $298(5)$ \\
\hline
\end{tabular}
\end{center}
\end{table}

\section{\label{sect:pion-corr}
  Pion correlation functions and the current renormalization}
The operators calculated on a lattice must be renormalized in order
to compare the results with other lattice studies and phenomenology.
In the case of the vector quark current, the renormalization constant is determined by the total 
charge measured as $g_V=F_1(0)$. 
The axial vector quark current renormalization constant $Z_A$ can be determined from 
the relation between  the local axial vector current $A_0$ and the true (partially conserved) 
axial vector current $\mathcal{A}_0$ associated with the axial transformation 
of the DW fermion integral \cite{blum2000}:
\begin{equation}\label{eqn:Za-ratio}
\langle\pi|\mathcal{A}_0|0\rangle  =  Z_A\langle\pi|A_0|0\rangle,
\quad \frac{\langle\mathcal{A}_0(t) \tilde{J_5}(0)\rangle}
        {\langle A_0(t) \tilde{J_5}(0)\rangle}
    \to Z_A,\,t\to\infty,
\end{equation}
where $\tilde{J}_5$ is the smeared pseudoscalar density operator.
Averaging  the ratio (\ref{eqn:Za-ratio}) over the plateau region $10\le t \le 54$, we extract
$Z_A$ with high precision as shown in Tab.~\ref{tab:pion-bare-quantities}.
We determine the pion mass, the residual mass $m_{res}$ and the pion decay constant from
the simultaneous fit of the following correlators of local operators:
\begin{align*}
\langle A_0(t) \tilde{J}_5(0)\rangle &= 
  c_{smear} A_5
  \left(e^{-m_\pi t} - e^{-m_\pi (L_t-t)}\right),\\
\langle J_5(t) \tilde{J}_5(0)\rangle &= 
  c_{smear} B_5
  \left(e^{-m_\pi t} + e^{-m_\pi (L_t-t)}\right),\\
\langle J_{5q}(t) \tilde{J}_5(0)\rangle &= 
  c_{smear} m_{res} B_5 
  \left(e^{-m_\pi t} + e^{-m_\pi(L_t-t)}\right),
\end{align*}
where $J_{5q}$ is the DW mid-point contribution to the divergence of the axial vector current
and $c_{smear}$ is the factor due to the source smearing, which is evaluated separately.
Constants 
$A_5 = f_\pi^2 m_\pi^2 / 4Z_A (m_q+m_{res})$
and 
$B_5 = f_\pi^2m_\pi^3 / 8(m_q+m_{res})^2$ 
provide us with two ways to extract the pion decay constant, $f_\pi$.
The obtained values agree within errors. 
Results are summarized in Tab.~\ref{tab:pion-bare-quantities}.

At the time of the talk, the lattice scale had only been set for the coarse lattice, using the
$\chi$PT extrapolated $\Omega$ baryon mass \cite{rbc-ukqcd-gauge}. 
Hence, we set the fine lattice scale by comparing the lattice values of 
the pion decay constant $af_\pi$ and the nucleon mass $am_N$
on the coarse lattice to that on the fine lattices, 
linearly interpolated in $(m_\pi/f_\pi)^2$ to point $x^* = \left.(m_\pi/f_\pi)^2\right|_{coarse}$:
\begin{equation*}
(af_\pi)^* / (af_\pi)^{coarse} = 0.7369(15),
\quad (am_N)^* / (am_N)^{coarse} = 0.7530(54)
\end{equation*}
Since the discrepancy between these ratios is smaller than the uncertainty in $a^{coarse}$,  we set $a^{fine} = 0.0841(14)$.

\begin{table}[ht]
\caption{\label{tab:pion-bare-quantities}
  Pseudoscalar meson quantities. 
  The residual mass is shown in Tab.~1.
  The fact that $Z_A g_V$ is so close to unity shows the close agreement between  
  the vector and axial current renormalization constants.
  }
\begin{center}\begin{tabular}{l|llllll}
\hline
$a \mathrm{ [fm]}$  & $am_l/am_h$  & $am_\pi$  & $af_\pi$  & $Z_A$ & $Z_A g_V$ & $am_N$\\
\hline
0.114 & 
0.005/0.04 & 0.1900(1) & 0.08615(13) & 0.71722(4) & & \\
\hline
0.086 & 
0.008/0.03 & 0.1729(1) & 0.06707(11) & 0.74530(4) & 0.988(4) & 0.5338(25) \\
& 
0.006/0.03 & 0.1516(1) & 0.06460(8)  & 0.74523(3) & 0.999(4) & 0.5048(24) \\
& 
0.004/0.03 & 0.1269(1) & 0.06229(11) & 0.74494(4) & 1.000(4) & 0.4758(12) \\
\hline
\end{tabular}\end{center}
\end{table}

\begin{figure}
\begin{minipage}{.49\textwidth}
\begin{center}
\includegraphics[angle=-90,width=\textwidth]{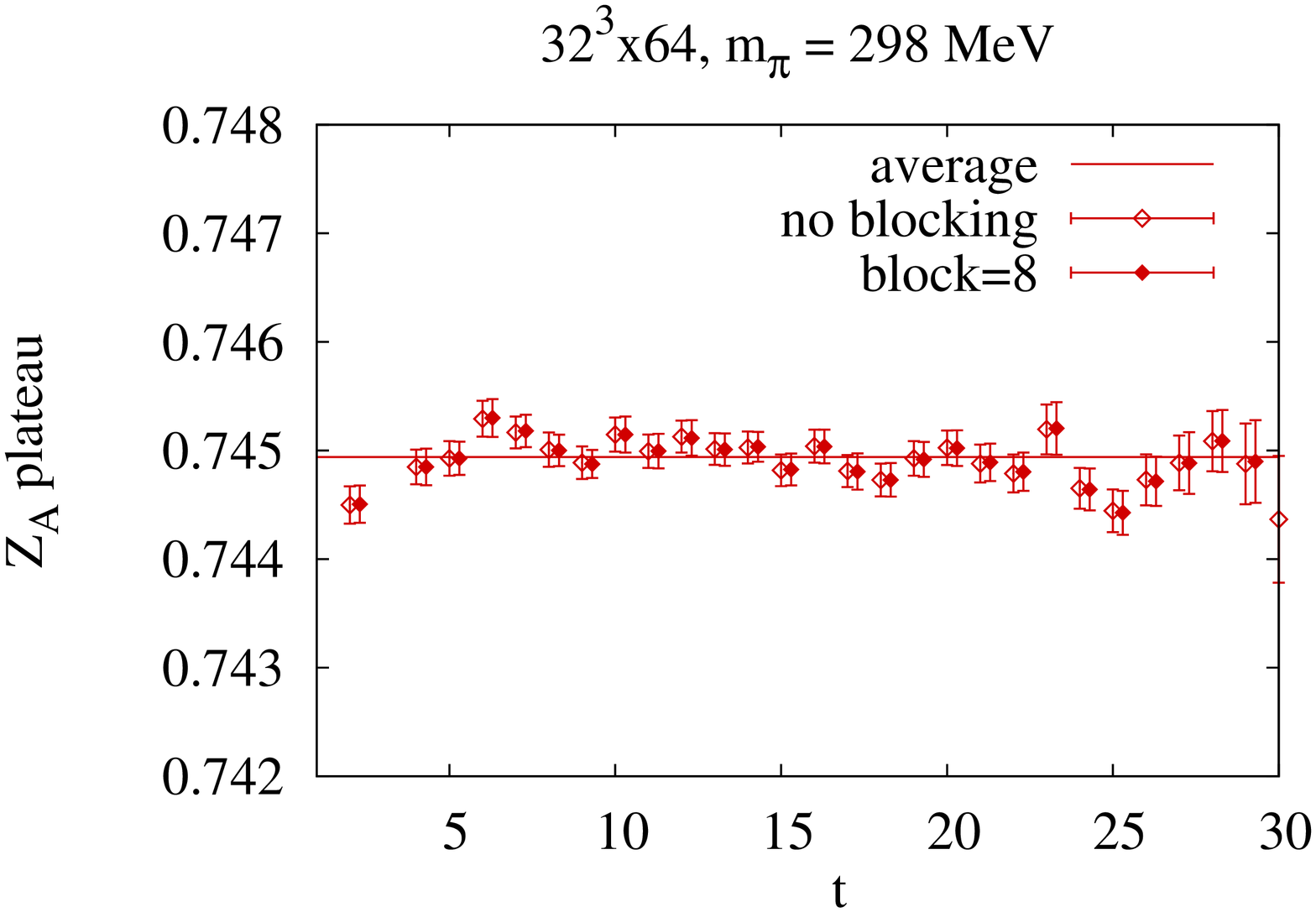}
\vspace{3pt}\\
\vspace{-10pt}
(a)
\end{center}
\end{minipage}
\begin{minipage}{.49\textwidth}
\begin{center}
\includegraphics[width=\textwidth]{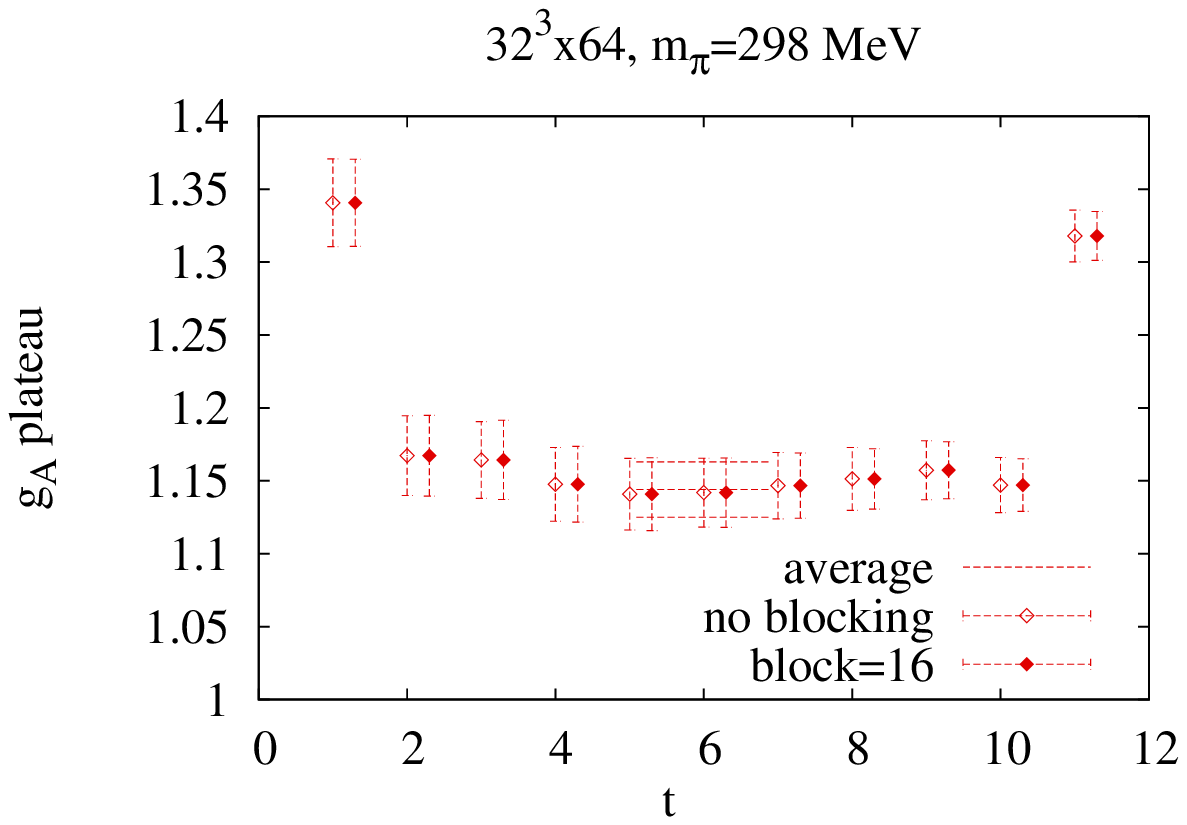}\\
\vspace{-10pt}
(b)
\end{center}
\end{minipage}
\caption{\label{fig:Za-pltx} Ratio determining the renormalization constant $Z_A$ (a) 
  and the plateau for the   axial charge $g_A$ (b) for $m_\pi=298\,\mathrm{MeV}$. 
  Both graphs show the average and the points with/without
    Jackknife binning.}
\end{figure}
  
\section{\label{sect:ff-results}
  Nucleon form factors}
To extract the quark current matrix elements between  nucleon states, we use the standard ratio of the
momentum projected correlation functions \cite{milc-ff-papers}:
\begin{equation*}
R^{\mathcal{O}}(T,\tau;P^\prime, P) = 
  C^{\mathcal{O}}_{3pt}\cdot
  \left(\begin{array}{c}C_{2pt} \text{ combination}\\\text{at } T, \tau, T-\tau\end{array}\right)
  \to \langle P^\prime|\mathcal{O}|P\rangle, 
\quad\text{with } T\to\infty , 
\end{equation*}
where the quantity in brackets represents  the appropriate combination of 
two-point functions to cancel out normalization factors at the source and sink. 
We calculate the following operators, which we use to extract the
corresponding (generalized) form factors
\begin{gather*}
\langle P^\prime|\bar q \gamma^\mu q|P\rangle = 
  \bar{U}(P^\prime)\left[F_1(Q^2)\gamma^\mu + F_2(Q^2)\frac{\sigma^{\mu\nu}q_\nu}{2m_N}\right]U(P),\\
\langle P^\prime|\bar q \gamma^\mu\gamma^5 q|P\rangle = 
  \bar{U}(P^\prime)\left[G_A(Q^2)\gamma^\mu\gamma^5 + G_P(Q^2)\frac{q^\mu}{2m_N}\right]U(P),\\
\langle P^\prime|\mathcal{O}^{\mu_1\dotsm\mu_n}_{[\gamma^5]}|P\rangle = 
  \bar{U}(P^\prime)\left[\overset{(\sim)}{A}_{n0}(Q^2)
                       \gamma^{\{\mu_1} [\gamma^5] 
                       \bar{P}^{\mu_2}\dotsm \bar{P}^{\mu_n\}} + \dotsb
                   \right] U(P),\\
\text{where\ \ } \mathcal{O}_{[\gamma^5]}^{\mu_1\dotsm\mu_n} = 
  \bar{q}\gamma^{\{\mu_1} [\gamma^5]
         i\overleftrightarrow{D}^{\mu_2}\dotsm
         i\overleftrightarrow{D}^{\mu_n\}} q,
  \quad q^\mu = {P^\prime}^\mu - P^\mu,
  \quad \bar{P}^\mu = \frac12 \left({P^\prime}^\mu + P^\mu\right).
\end{gather*}

\begin{figure}[th]
\begin{center}
\begin{minipage}{.49\textwidth}
\begin{center}\includegraphics[width=\textwidth]{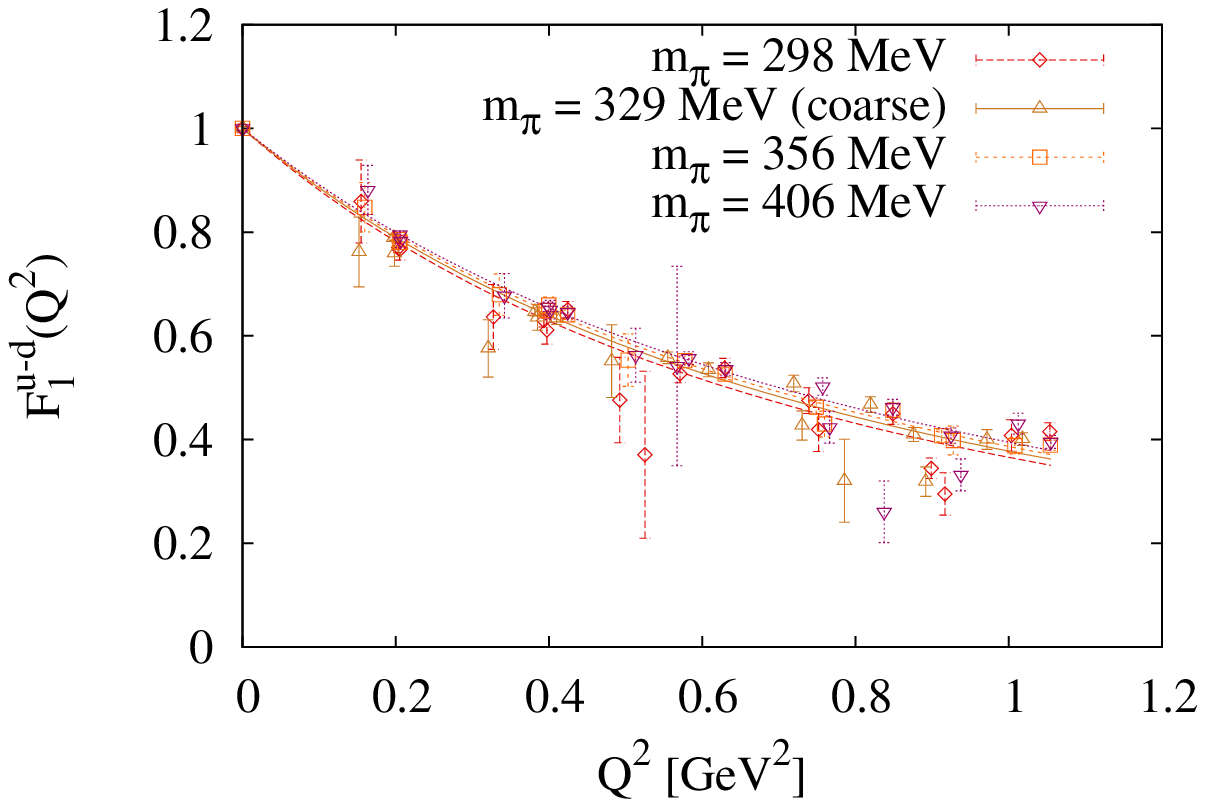}\end{center}
\end{minipage}
\begin{minipage}{.49\textwidth}
\begin{center}
\includegraphics[width=\textwidth]{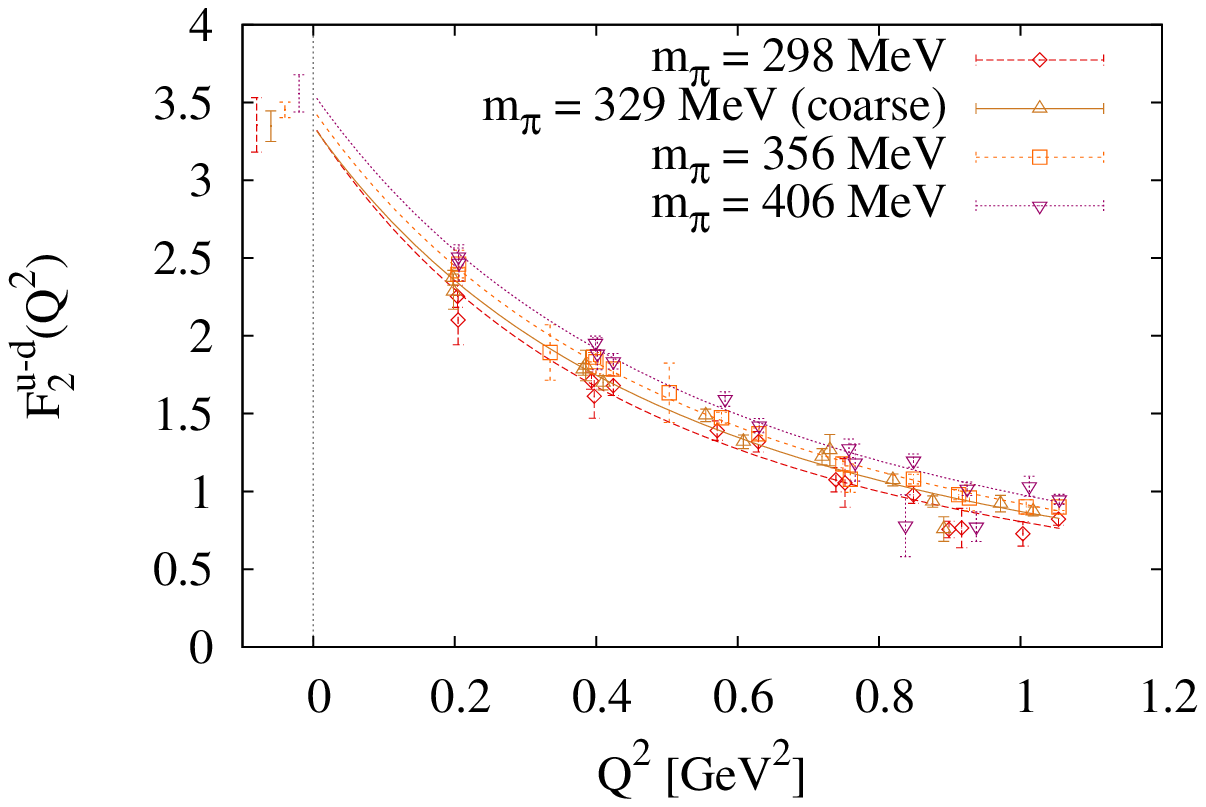}\end{center}
\end{minipage}
\end{center}
\vspace{-10pt}
\caption{\label{fig:vector-formfactor} 
  The vector form factors $F_1^{u-d}(Q^2)$ and $F_2^{u-d}(Q^2)$ for three different $m_\pi$.
  The curves correspond to the dipole fit in range $0 \le Q^2 \le 0.4\,\mathrm{GeV^2}$ 
  and range $0.2 \le Q^2 \le 0.8\,\mathrm{GeV^2}$, respectively.
  On the $F_2^{u-d}(Q^2)$ plot, the errorbars to the left of $Q^2=0$ line show the uncertainty
  in the determination of $F_2^{u-d}(0)$.
  }
\end{figure}

On Fig.~\ref{fig:vector-formfactor}, we show the results for the isovector form factors of the 
the vector current $F_1^{u-d}(Q^2)$ and $F_2^{u-d}(Q^2)$, where
both form factors are fitted with the dipole formula. 
The form factors $F_1(Q^2)$ and $F_2(Q^2)$
are renormalized with $g_V=F_{1}^{bare}(0)$, so that $F_1(0) \equiv 1$. 

These initial results for both form factors,
using only a small fraction of the full set of planned domain wall ensembles,
are already of high quality and consistent on coarse and fine lattices.
As expected, decreasing the pion mass leads to a larger Dirac mean squared radius and
correspondingly to a steeper form factor $F_1(Q^2)$.
Since the intermediate pion mass is nearly halfway between the light and heavy masses, 
and the form factors differ only by a very small amount, 
expanding it to leading order one would expect, 
in the absence of any lattice spacing dependence, 
that the intermediate curve would also be halfway between the upper and lower curve.  
The fact that the coarse lattice result at the intermediate mass indeed lies halfway 
between the two fine lattice results is a clear signature that lattice artifacts 
associated with the lattice spacing are very small for these form factors.

\begin{figure}[th]
\begin{center}
\begin{minipage}{.49\textwidth}
\begin{center}
\includegraphics[width=\textwidth]{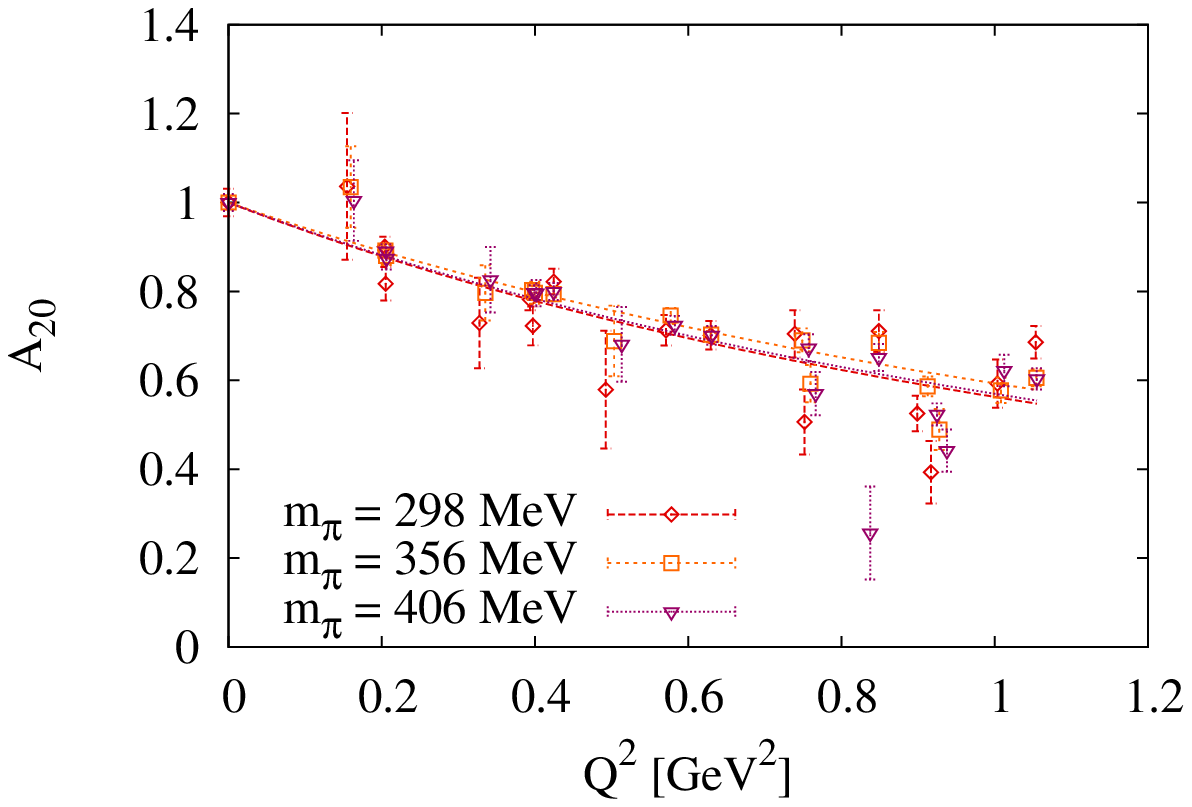}\\
\vspace{-10pt}
\includegraphics[width=\textwidth]{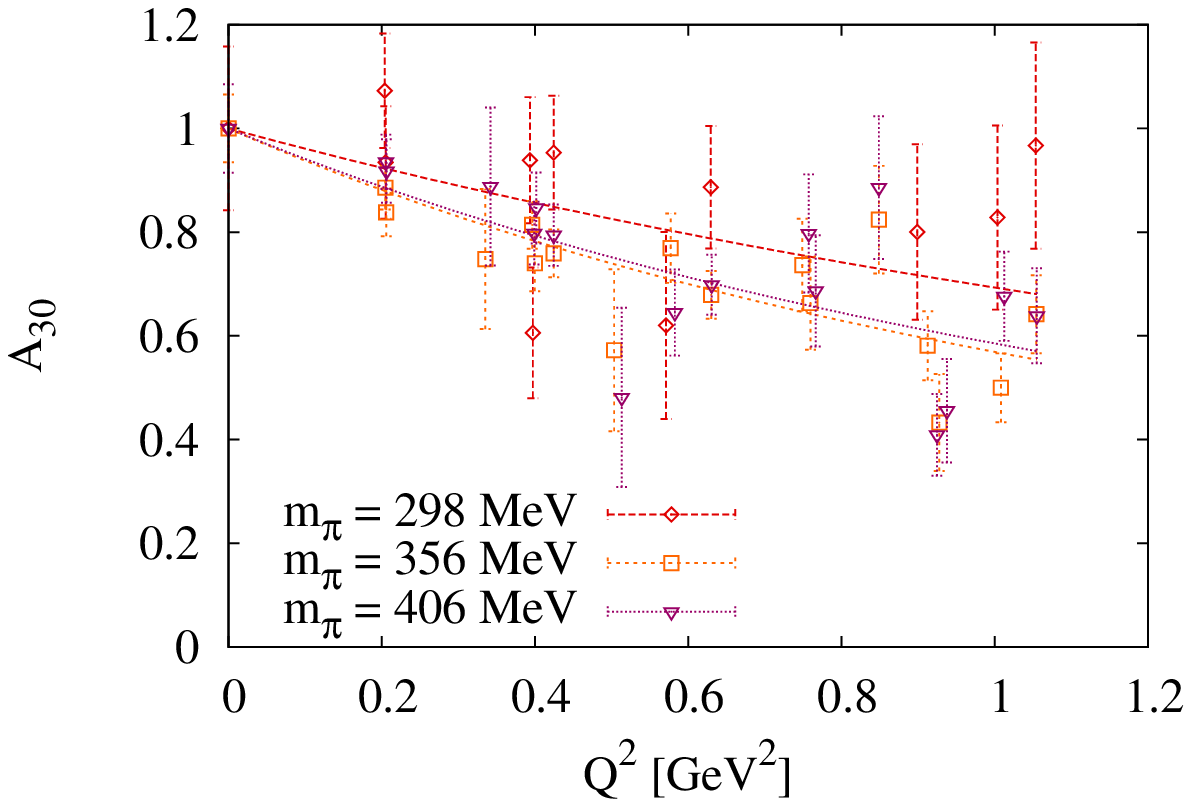}
\end{center}
\vspace{-10pt}
\centerline{(a)}
\end{minipage}
\begin{minipage}{.49\textwidth}
\begin{center}
\includegraphics[width=\textwidth]{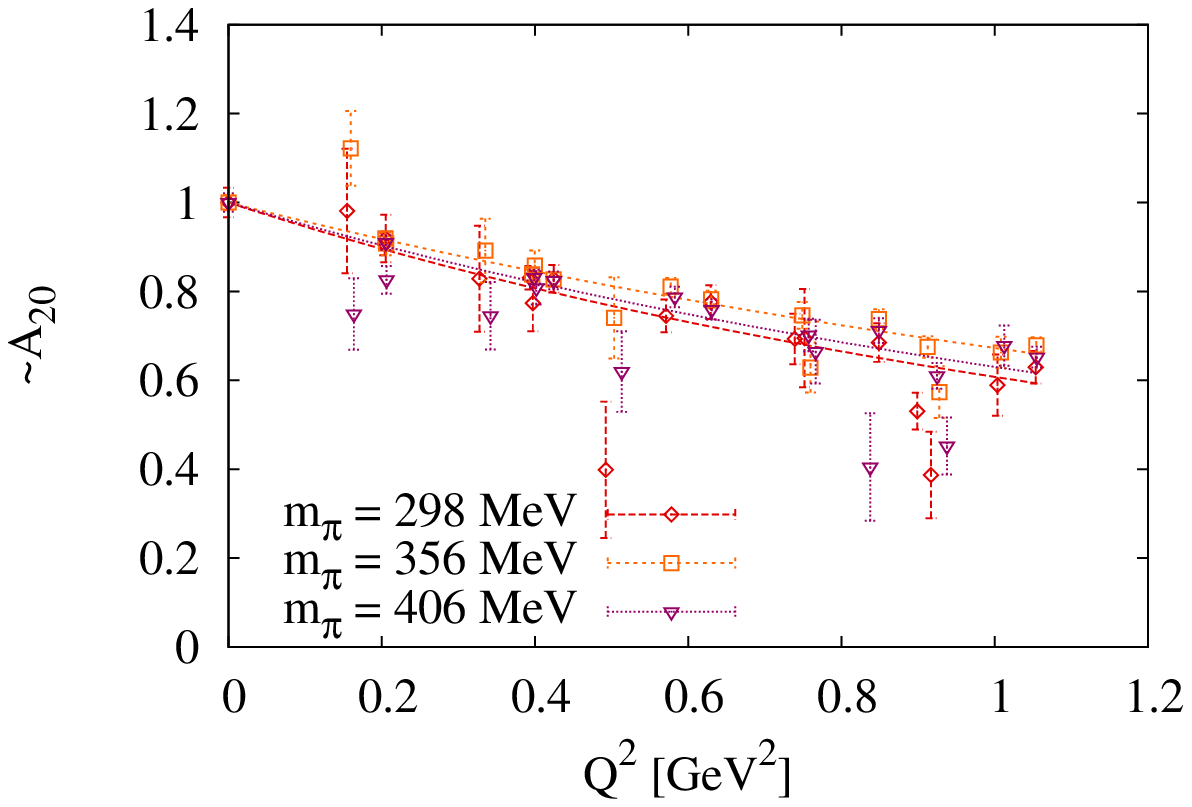}\\
\vspace{-10pt}
\includegraphics[width=\textwidth]{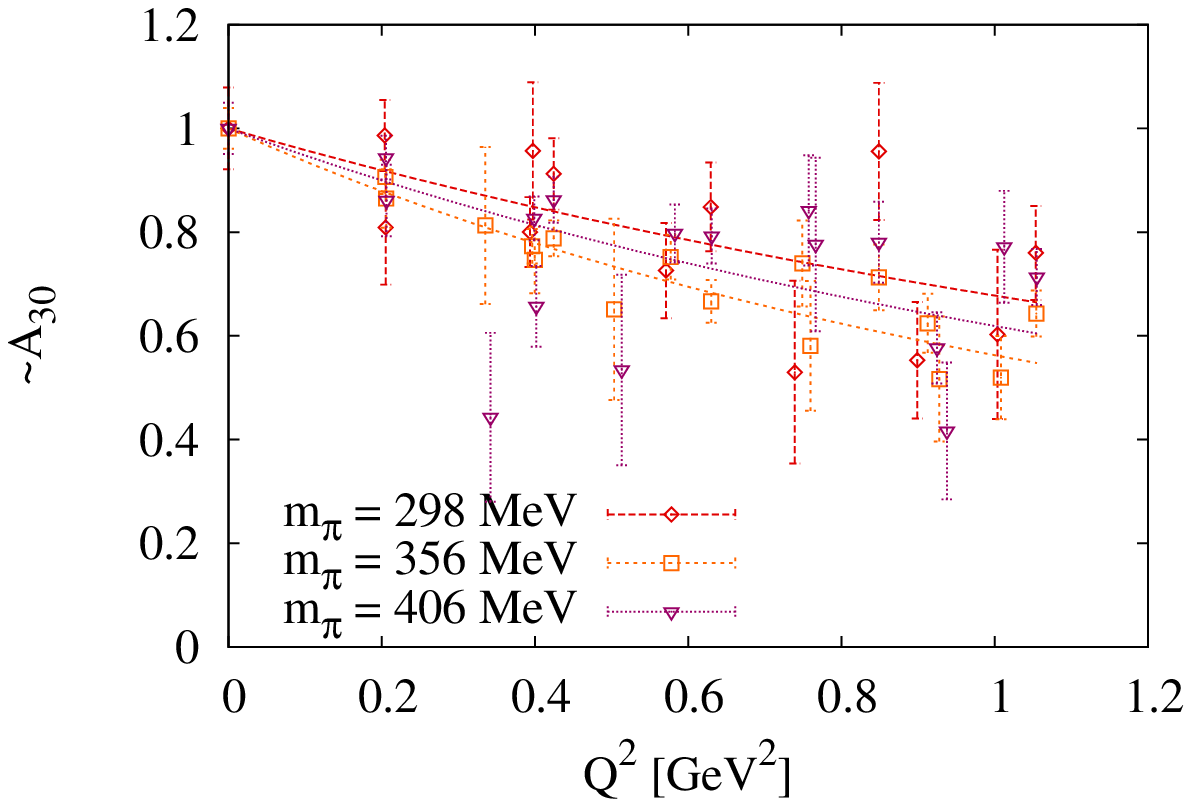}
\end{center}
\vspace{-10pt}
\centerline{(b)}
\end{minipage}
\end{center}
\vspace{-10pt}
\caption{\label{fig:gen-ff}
  Helicity-even (a) and helicity-odd (b) generalized form factors on fine lattices.
  All the form factors are normalized to one at $Q^2=0$.
  Lines represent one-parameter dipole fits.}
\end{figure}

The only generalized form factors whose dependency on the transferred momentum $Q^2$ can be
extracted reliably with the present statistics are the leading ones, $A_{n0}$ and $\tilde{A}_{n0}$. 
In Fig.~\ref{fig:gen-ff} we show the results for these form factors, normalized to unity at the
zero momentum transfer $Q^2=0$.  As one goes to higher moments, involving correspondingly more 
derivatives in the twist-two operators, the statistical errors increase as expected. 
However, when the statistics are eventually increased by up to an order of magnitude, 
these higher generalized form factors will also be well determined.

\section{\label{sect:conclusions}
  Conclusions}
  
By virtue of performing eight measurements per fine domain wall lattice configuration, 
our initial calculations on ensembles of roughly 300 to 500 configurations 
show a good statistical precision
for nucleon generalized form factors.
Comparison of results with coarse and fine lattices indicates that the errors in 
the form factors arising from lattice spacing artifacts are quite small for domain wall fermions. 
The study of other generalized form factors, such as those related to the quark angular momentum, 
will require calculation of renormalization constants for the corresponding operators, 
which is in progress.

\section*{Acknowledgements}

This work was supported in part by U.S. DOE Contract No. DE-AC05-06OR23177 under which JSA 
operates Jefferson Laboratory, by the DOE Office of Nuclear Physics under grants 
DE-FG02-94ER40818, 
DE-FG02-04ER41302, 
DE-FG02-96ER40965, 
by the DFG (Forschergruppe Gitter-Hadronen-Ph\"anomenologie), 
and  the EU Integrated Infrastructure Initiative Hadron Physics (I3HP) under contract  
RII3-CT-2004-506078. 
W.S. acknowledges support by the National Science Council of Taiwan under grants
NSC96-2112-M002-020-MY3 and NSC96-2811-M002-026,
K.O. acknowledges support from the  Jeffress  Memorial Trust grant J-813,
Ph. H. and B. M. acknowledge support by the Emmy-Noether program of the DFG, and  Ph. H., M.P. 
and W. S. acknowledge support by the A.v. Humboldt-foundation through the Feodor-Lynen program.
Ph.~H. thanks the Excellence Cluster Universe at the TU Munich for support.
This research used resources under the INCITE and ESP programs of the Argonne Leadership 
Computing Facility at Argonne National Laboratory, which is supported by the Office of Science 
of the U.S. Department of Energy under contract DE-AC02-06CH11357, resources provided by the DOE 
through the USQCD project at Jefferson Lab and through its support of the MIT Blue Gene/L
under grant DE-FG02-05ER25681,  
resources provided by the William and Mary Cyclades Cluster, and resources provided by the
New Mexico Computing Applications Center (NMCAC) on Encanto.  We are indebted to members of 
the MILC, RBC, and UKQCD Collaborations for providing the dynamical quark configurations that 
made our full QCD calculations possible.

\end{document}